\pdfoutput=1

\documentclass[11pt,a4paper]{article}
\usepackage[hyperref]{acl2020-2}
\usepackage[utf8]{inputenc}
\usepackage{times}
\usepackage{latexsym}

\usepackage{url}
\usepackage{enumitem}
\usepackage{microtype}
\usepackage{booktabs}
\usepackage{multirow}
\usepackage{array,graphicx}
\usepackage{pifont}
\usepackage{multicol}
\usepackage{makecell}
\usepackage{rotating}
\usepackage{amssymb}
\usepackage{txfonts}
\usepackage{subfig}
\usepackage{tikz}
\usepackage{pgfplots}
\pgfplotsset{compat=1.15}
\usepackage{stmaryrd} 

\usepackage[normalem]{ulem}
\makeatletter
\def\squiggly{\bgroup \markoverwith{\textcolor{red}{\lower3.5\p@\hbox{\sixly \char58}}}\ULon}
\makeatother

\colorlet{good}{green!40!black}
\colorlet{bad}{red!50!black}

\title{Keyphrase Generation for Scientific Document Retrieval}

\author{Florian Boudin$^1$, Ygor Gallina$^1$, Akiko Aizawa$^2$\\
  $^{1}$LS2N, Université de Nantes, France \quad
  \texttt{first.last@univ-nantes.fr} \\ 
  $^{2}$National Institute of Informatics, Tokyo, Japan \quad
  \texttt{aizawa@nii.ac.jp}\\
  }

\date{}

\begin{document}
\maketitle
\begin{abstract}
  Sequence-to-sequence models have lead to significant progress in keyphrase generation, but it remains unknown whether they are reliable enough to be beneficial for document retrieval.
  This study provides empirical evidence that such models can significantly improve retrieval performance, and introduces a new extrinsic evaluation framework that allows for a better understanding of the limitations of keyphrase generation models.
  Using this framework, we point out and discuss the difficulties encountered with supplementing documents with --not present in text-- keyphrases, and generalizing models across domains.
  Our code is available at \url{https://github.com/boudinfl/ir-using-kg}.
\end{abstract}

\vspace{-0.9em}

\section{Introduction}

\vspace{-0.3em}
With the exponential growth of the scientific literature~\cite{lutz-rudiger2015}, retrieving relevant scientific papers becomes increasingly difficult.
Keywords, also referred to as keyphrases, provide an effective way to supplement paper indexing and improve retrieval effectiveness in scientific digital libraries~\cite{barker1972comparative,zhai-1997-fast,Gutwin:1999:IBD:338985.338996,doi:10.1002/asi.22985}.
However, only few documents have assigned keyphrases, and those who do were, for the most part, self-labeled by their authors, thus exhibiting annotation inconsistencies~\cite{strader2011author,suzuki-etal-2011-analyzing}.
This has motivated an active line of research on automatic keyphrase extraction (see~\citet{hasan-ng-2014-automatic} for an overview) and, more recently, keyphrase generation~\cite{meng-etal-2017-deep}, where the task is to find a set of words and phrases that represents the main content of a document.

Although models for predicting keyphrases have been extensively evaluated on their ability to reproduce author's keywords, it still remains unclear whether they can be usefully applied in information retrieval.
One reason for this lack of evidence may have been their relatively low performance discouraging attempts at using them for indexing~\cite{liu-etal-2010-automatic,hasan-ng-2014-automatic}.
Yet, recently proposed models not only achieve much better performance, but also display a property that may have a significant impact on retrieval effectiveness: the capacity to generate keyphrases that do not appear in the source text.
%
These \textit{absent} keyphrases do not just highlight the topics that are most relevant, but provide some form of semantic expansion by adding new content (e.g.~synonyms, semantically related terms) to the index~\cite{greulich2011scientific}.
The goal of this paper is two-fold: to gather empirical evidence as to whether current keyphrase generation models are good enough to improve scientific document retrieval, and to gain further insights into the performance of these models from an extrinsic perspective.
Our contributions are listed as follows:
%
%
\begin{itemize}
    \item We report significant improvements for strong retrieval models on a standard benchmark collection, showing that keyphrases produced by state-of-the-art models are consistently helpful for document retrieval, even, to our surprise, when author keywords are provided.

    \item We introduce a new extrinsic evaluation framework for keyphrase generation that allows for a deeper understanding of the limitations of current models. Using it, we discuss the difficulties associated with domain generalization and absent keyphrase prediction.
\end{itemize}

\section{Methodology}

This section presents our methodology for assessing the usefulness (\S\ref{subsec:extr_eval}) of keyphrase generation (\S\ref{subsec:key_gen}) in scientific document retrieval (\S\ref{subsec:doc_ret}).

\subsection{Scientific Document Retrieval}
\label{subsec:doc_ret}

Here, we focus on the task of searching through a collection of scientific papers for relevant documents.
All of our experiments are conducted on the NTCIR-2 test collection~\cite{ntcir-2} which is, to our knowledge, the only available benchmark dataset for that task.
It contains 322,058 documents\footnote{Scientific abstracts and summaries of research results.} (title and abstract pairs) and 49 search topics (queries) with relevance judgments.
Most of the documents (98.6\%) include author keywords (4.8 per doc.~on avg.), which we later use to investigate the performance of keyphrase generation models.

Documents cover a broad range of domains from pure science to social sciences and humanities, although half of the documents are about engineering and computer science.
Queries are also categorized into one or more research fields (e.g.~science, chemistry, engineering), the original intent being to help retrieval models in narrowing down the search space.
We follow common practice and use short\footnote{\texttt{<description>} field of topic description.} queries with binary relevance judgments (i.e.~without ``partially relevant'' documents).


We consider two standard \textit{ad-hoc} retrieval models to rank documents against queries: BM25 and query likelihood (QL), both implemented in the Anserini IR toolkit~\cite{Yang:2017:AEU:3077136.3080721}.
These models use unsupervised techniques based on corpus statistics for term weighting, and will therefore be straightforwardly affected when keyphrases are added to a document.
We further apply a pseudo-relevance feedback method, known as RM3~\cite{abdul2004umass}, on top of the models to achieve strong, near state-of-the-art retrieval results~\cite{Lin:2019:NHC:3308774.3308781,Yang:2019:CEH:3331184.3331340}.
For all models, we use Anserini's default parameters.

To verify the effectiveness of the adopted retrieval models, we compared their performance with that of the best participating systems in NTCIR-2.
Retrieval performance is measured using mean average precision (MAP) and precision at 10 retrieved documents (P@10).
MAP measures the overall ranking quality and P@10 reflects the number of relevant documents on the first page of search results.
Documents are indexed with author keywords, same as for participating systems.
Results are presented in Table~\ref{tab:ir_intrinsic}.
We see that the considered retrieval models achieve strong performance, even outperforming the best participating system by a substantial margin.
Note that the two best-performing systems use pseudo-relevance feedback, and that the second-ranked system is based on BM25.

\begin{table}[!ht]
    \centering
    \begin{tabular}{l|cc}
        \toprule
            \textbf{Model} & 
            \textbf{MAP} & 
            \textbf{P@10} \\
        \midrule
            
            BM25+RM3 &
            \textbf{35.17} & \textbf{38.57} \\ 
            
            QL+RM3 &
            33.00 & 34.90 \\ 
            
            1\textsuperscript{st} {\small \cite{fujita2001notes}} & 
            31.93 & 37.35 \\ 
            
            BM25 &
            31.38 & 36.33 \\ 
            
            2\textsuperscript{nd} {\small \cite{murata2001crl}} & 
            31.31 & 36.12 \\ 
            
            QL &
            30.63 & 34.08 \\ 
            
            3\textsuperscript{rd} {\small \cite{chen2001berkeley}} &
            26.24 & 33.88 \\ 
            
        \bottomrule
    \end{tabular}
    \caption{Retrieval effectiveness of the considered models and the best participating systems on NTCIR-2.}
    \label{tab:ir_intrinsic}
\end{table}

\subsection{Keyphrase Generation}
\label{subsec:key_gen}

Keyphrase generation is the task of producing a set of words and phrases that best summarise a document~\cite{evans-zhai:1996:ACL}.
In contrast with most previous work that formulates this task as an extraction problem  (a.k.a.~keyphrase extraction), which can be seen as ranking phrases extracted from a document, recent neural models for keyphrase generation are based on sequence-to-sequence learning~\cite{NIPS2014_5346,bahdanau2014neural}, thus potentially allowing them to generate any phrase, also beyond those that appear verbatim in the text.
In this study, we consider the following two neural keyphrase generation models:
\begin{description}[]
    \item[seq2seq+copy] \cite{meng-etal-2017-deep} is a sequence-to-sequence model with attention, augmented with a copying mechanism~\cite{gu-EtAl:2016:P16-1} to predict phrases that rarely occur. The model is trained with document-keyphrase pairs and uses beam search decoding for inference.

    \item[seq2seq+corr] \cite{chen-etal-2018-keyphrase} extends the aforementioned model with correlation constraints. It employs a coverage mechanism~\cite{tu-etal-2016-modeling} that diversifies attention distributions to increase topic coverage, and a review mechanism to avoid generating duplicates.
\end{description}

We implemented the models in PyTorch~\cite{paszke2017automatic} using  AllenNLP~\cite{gardner-etal-2018-allennlp}.
Models are trained on the KP20k dataset~\cite{meng-etal-2017-deep}, which contains 567,830 scientific abstracts with gold-standard, author-assigned keywords (5.3 per doc.~on avg.). 
We use the parameters suggested by the authors for each model.

To validate the effectiveness of our implementations, we conducted an intrinsic evaluation by counting the number of exact matches between predicted and gold keyphrases.
We adopt the standard metric and compute the f-measure at top 5, as it corresponds to the average number of keyphrases in KP20k and NTCIR-2, that is, 5.3 and 4.8, respectively.
We also examine cross-domain generalization using the KPTimes news dataset~\cite{gallina-boudin-daille-2019-kptimes}, and include a state-of-the-art unsupervised keyphrase extraction model~\cite[henceforth mp-rank]{boudin-2018-unsupervised} for comparison purposes.
This latter baseline also provides an additional relevance signal based on graph-based ranking whose usefulness in retrieval will be tested in subsequent experiments.
Results are reported in Table~\ref{tab:kg_intrinsic}.
Overall, our results are consistent with those reported in~\cite{meng-etal-2017-deep,chen-etal-2018-keyphrase}, demonstrating the superiority of well-trained neural models over unsupervised ones, and stressing their lack of robustness across domains.
Rather surprisingly, seq2seq+corr is outperformed by seq2seq+copy which indicates that relevant, yet possibly redundant, keyphrases are filtered out by the added mechanisms for promoting diversity in the output.

\begin{table}[ht!]
    \centering
    \begin{tabular}{l|ccc}
        \toprule
            \textbf{Model} & 
            \textbf{~KP20k~} & 
            \textbf{NTCIR-2} &
            \textbf{KPTimes} \\
        \midrule
        s2s+copy & \textbf{27.75} & \textbf{23.90} & \textbf{16.47}  \\
        
        s2s+corr & 23.78 & 22.27 & 11.73 \\
        
        mp-rank  & 14.67 & 18.10 & 14.59  \\
        \bottomrule
    \end{tabular}
    \caption{f-measure at top-$5$ predicted keyphrases. Stemming is applied to reduce the number of mismatches.}
    \label{tab:kg_intrinsic}
\end{table}

\subsection{Extrinsic Evaluation Framework}
\label{subsec:extr_eval}

Our goal is to find out whether the keyphrase generation models described above are reliable enough to be beneficial for document retrieval.
To do so, we contrast the performance of the retrieval models with and without automatically predicted keyphrases.
Two initial indexing configurations are also examined: title and abstract only ($T$+$A$), and title, abstract and author keywords ($T$+$A$+$K$).
The idea here is to investigate whether generated keyphrases simply act as a proxy for author keywords, or instead supplement them. 

Unless mentioned otherwise, the top-5 predicted keyphrases are used to expand documents, which is in accordance with the average number of author keywords in NTCIR-2.
We evaluate retrieval performance in terms of MAP and omit P@10 for brevity.
We use the Student’s paired t-test to assess statistical significance of our retrieval results at $p<0.05$~\cite{Smucker:2007:CSS:1321440.1321528}.

\section{Results}

Results for retrieval models using keyphrase generation are reported in Table~\ref{tab:ir_results}.
We note that indexing keyphrases generated by seq2seq+copy, which performs best in our intrinsic evaluation, significantly improves retrieval effectiveness for all models.
More interestingly, gains in effectiveness are also significant when both keyphrases and author keywords are indexed, indicating they complement each other well.
This important finding suggests that predicted keyphrases are consistently helpful for document retrieval, and should be used even when author keywords are provided.
Another important observation is that while both keyphrase generation models perform reasonably well in our intrinsic evaluation on NTCIR-2 (cf.~Table~\ref{tab:kg_intrinsic}, column 3), their impact on retrieval effectiveness are quite different, as only s2s+copy reaches consistent significance.
This finding advocates for the importance of using document retrieval as an extrinsic evaluation task for keyphrase generation.

\newcommand*\sign[1]{\phantom{$^\dagger$}#1$^\dagger$}
\newcommand*\signn[1]{\phantom{$^\ddagger$}#1$^\ddagger$}

\begin{table}[!ht]
    \centering
    \setlength{\tabcolsep}{4pt}
    \resizebox{.485\textwidth}{!}{%
    \begin{tabular}{l|cccc}
        \toprule
            \textbf{Index} & 
            \textbf{BM25} & 
            \textbf{+RM3} & 
            \textbf{QL} & 
            \textbf{+RM3}  \\
        \midrule
        $T$+$A$  & 29.16 & 31.93 & 28.98 & 31.47 \\
        \quad+ s2s+copy  & \sign{30.54} & \sign{\textbf{34.30}} & \sign{30.58} & \sign{33.26} \\
        \quad+ s2s+corr  & \sign{30.30} & 33.24 & 29.76 & 31.38 \\
        \quad+ mp-rank  & 29.24 & 32.27 & 29.57 & 32.29 \\
        \midrule
        $T$+$A$+$K$ & 31.38 & 35.17 & 30.63 & 33.00 \\
        \quad+ s2s+copy & 31.55 & \signn{\textbf{36.53}} & \signn{31.70} & \signn{35.15} \\
        \quad+ s2s+corr & 31.37 & 35.84 & 31.14 & 33.65 \\
        \quad+ mp-rank & 31.38 & 35.18 & 31.23 & 33.47 \\
        \bottomrule
    \end{tabular}
    }
    \caption{MAP scores for retrieval models using various indexing configurations. 
    $\dagger$ and $\ddagger$ indicate significance over $T$+$A$ and $T$+$A$+$K$, respectively.}
    \label{tab:ir_results}
\end{table}

Overall, BM25+RM3 achieves the best retrieval effectiveness, confirming previous findings on \textit{ad-hoc} retrieval in limited data scenarios~\cite{Lin:2019:NHC:3308774.3308781}.
For clarity and conciseness, we focus on this model in the rest of this paper.
Encouraging diversity in keyphrases seems not to be appropriate for retrieval, as  seq2seq+corr consistently gives lower results than seq2seq+copy.
It is also interesting to see that the effectiveness gains of query expansion (RM3) and document expansion are additive, suggesting that they provide different but complementary relevance signals.
Moreover, our results show that query expansion is more effective, which is in line with past work~\cite{billerbeck2005document}.

\begin{figure}[!ht]
    \centering
    \subfloat{
    \hspace{1em}
    \begin{tikzpicture}[trim axis left,trim axis right]
    \begin{axis}[
        width=0.54\textwidth,
        height=3.8cm,
	    ymin=31.5,
	    ymax=35.0,
	    xmin=-1,
	    xmax=10,
	    xtick = {0, 1, 2, 3, 4, 5, 6, 7, 8, 9},
	    xticklabels={,,,},
	    ytick = {32, 33, 34},
	    tick pos=left,
	    grid style={dashed, gray!50},
	    xmajorgrids,
        ]
    
    \node[] at (axis cs: .09, 34.5) {{\small $T$+$A$}};
        
    \addplot[only marks, mark=square, color=black, thick, mark size=3pt] plot coordinates {
        (3, 33.46) 
        (4, 34.16) 
        (5, 34.30) 
        (6, 34.19) 
        (7, 33.43) 
        (8, 33.81) 
        (9, 33.96) 
    };
    
    \addplot[smooth, mark=*, blue] plot coordinates {
        (0, 31.93)
        (1, 32.61) 
        (2, 32.87) 
        (3, 33.46) 
        (4, 34.16) 
        (5, 34.30) 
        (6, 34.19) 
        (7, 33.43) 
        (8, 33.81) 
        (9, 33.96) 
    };
    
    \addplot[smooth, mark=o, red] plot coordinates {
        (0, 31.93)
        (1, 32.60) 
        (2, 32.64) 
        (3, 32.34) 
        (4, 32.90) 
        (5, 33.24) 
        (6, 32.93) 
        (7, 32.85) 
        (8, 33.11) 
        (9, 33.49) 
    };
    
    \addplot[smooth, mark=x, green!70!black] plot coordinates {
        (0, 31.93)
        (1, 32.10) 
        (2, 31.93) 
        (3, 32.26) 
        (4, 32.49) 
        (5, 32.06) 
        (6, 32.45) 
        (7, 32.59) 
        (8, 33.17) 
        (9, 33.14) 
    };
    
    \addplot[mark=none, black, dashed] coordinates {(0, 31.93) (9, 31.93)};
    
    \end{axis}
    \end{tikzpicture}
    }
    \vspace*{-1.2em}
    \subfloat{
    \hspace{1em}
    \begin{tikzpicture}[trim axis left,trim axis right]
    \begin{axis}[
        width=0.54\textwidth,
        height=3.8cm,
	    ymin=34.5,
	    ymax=38,
	    xmin=-1,
	    xmax=10,
	    xtick = {0, 1, 2, 3, 4, 5, 6, 7, 8, 9},
	    xticklabels={$N$=0\quad~~,1, 2, 3, 4, 5, 6, 7, 8, 9},
	    ytick = {35, 36, 37},
	    tick pos=left,
	    grid style={dashed, gray!50},
	    xmajorgrids,
	    legend style={font=\small},
	    legend entries={s2s+copy,s2s+corr,mp${-}$rank},
	    legend style={at={(0.95,1.01)}, anchor=north east, draw=none, fill=none, text width=2em,text height=1ex,text depth=.5ex}
        ]
    
    \node[] at (axis cs: .4, 37.5) {{\small $T$+$A$+$K$}};
    
    \addplot[smooth, mark=*, color=blue] plot coordinates {
        (0, 35.17)
        (1, 35.82) 
        (2, 35.71) 
        (3, 36.24) 
        (4, 36.85) 
        (5, 36.53) 
        (6, 36.43) 
        (7, 36.16) 
        (8, 35.48) 
        (9, 35.60) 
    };

    \addplot[smooth, mark=o, color=red] plot coordinates {
        (0, 35.17)
        (1, 35.12)
        (2, 34.94)
        (3, 35.53)
        (4, 35.33)
        (5, 35.84)
        (6, 35.92)
        (7, 35.79)
        (8, 35.38)
        (9, 35.74)
    };
    
    \addplot[smooth, mark=x, green!70!black] plot coordinates {
        (0, 35.17)
        (1, 34.84)
        (2, 34.85)
        (3, 35.44)
        (4, 35.25)
        (5, 35.18)
        (6, 34.95)
        (7, 34.85)
        (8, 35.04)
        (9, 35.42)
    };
    
    \addplot[only marks, mark=square, color=black, thick, mark size=3pt] plot coordinates {
        (3, 36.24) 
        (4, 36.85) 
        (5, 36.53) 
        (6, 36.43) 
    };
    
    \addplot[mark=none, black, dashed] coordinates {(0, 35.17) (9, 35.17)};

    \end{axis}
    
    \end{tikzpicture}
    }
    \caption{MAP scores for BM25+RM3 w.r.t. the number $N$ of predicted keyphrases. $\square$ denotes significance.}
    \label{fig:n_vs_perf}
\end{figure}
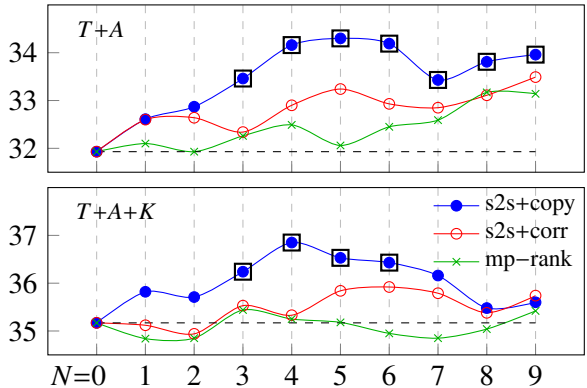

One hyper-parameter that we have deliberately left untouched so far is the number $N$ of predicted keyphrases that directly controls the precision-recall trade-off of keyphrase generation models.
To understand how this parameter affects retrieval effectiveness, we repeated our experiments by varying $N$ within the range $[0,9]$, and plotted the results in Figure~\ref{fig:n_vs_perf}.
Without author keywords, we observe that all models achieve gains, but only seq2seq+copy does yield significant improvements.
With author keywords, seq2seq+copy is again the only model that achieves significance, 
while the others show mixed results, sometimes even degrading scores.
One likely explanation for this is that these models produce keyphrases that cause documents to drift away from their original meaning.
We note that results are close to optimal for $N=5$, supporting our initial setting for this parameter.

From our experiments, it appears that unsupervised keyphrase extraction is not effective enough to significantly improve retrieval effectiveness.
The fact that keyphrase generation does so, suggests that the ability to predict absent keyphrases may be what enables better performance.
Yet counter-intuitively, we found that most of the gains in retrieval effectiveness are due to the high extractive accuracy of keyphrase generation models.
Results in Table~\ref{tab:all-abs-pres} show that expanding documents with only absent keyphrases is at best useless and at worst harmful, while using only present keyphrases brings significant improvements.
We draw two conclusions from this.
First, absent keyphrases may not be useful in practice unless they are tied to some form of domain terminology to prevent semantic drift.
Second, as generation does not yield improvements, keyphrase extraction models may be worth further investigation.
In particular, supervised models could theoretically provide similar results while being easier to train.

\begin{table}[ht!]
    \centering
    \setlength{\tabcolsep}{5pt}
    \resizebox{.485\textwidth}{!}{%
    \begin{tabular}{l|cc|cc}
        \toprule
        \multirow{2}{*}{\textbf{Model}} & \multicolumn{2}{c|}{$T$+$A$ {\small(cf.~31.93)}} & \multicolumn{2}{c}{$T$+$A$+$K$ {\small(cf.~35.17)}}\\
        ~ & \textbf{pres.} & \textbf{abs.} & \textbf{pres.} & \textbf{abs.} \\
        \midrule
        s2s+copy & 
            \sign{\textcolor{good}{\textbf{34.17}}} & ~\textcolor{good}{32.14}~ & 
            \sign{\textcolor{good}{\textbf{36.30}}} & ~\textcolor{bad}{34.97}~ \\
        s2s+corr &
            \textcolor{good}{32.97} & \textcolor{good}{31.96} &
            \textcolor{good}{36.09} & \textcolor{bad}{34.77} \\
        \bottomrule
    \end{tabular}
    }
    \caption{MAP scores for BM25+RM3 using the top-5 present or absent keyphrases. $\dagger$ indicates significance over indexing without predicted keyphrases.}
    \label{tab:all-abs-pres}
\end{table}

Neural models for keyphrase generation exhibit a limited generalization ability, which means that their performance degrades on documents that differ from the ones encountered during training (cf.~Table~\ref{tab:kg_intrinsic}, columns 3 and 4).
To quantify how much this affects retrieval effectiveness, we divided the queries into two disjoint sets: \textit{in-domain} for those that belong to research fields present in KP20k, and \textit{out-domain} for the others.
Results are presented in Table~\ref{tab:per_domain}.
The first thing we notice is the overall lower performance of \textit{out-domain} queries, which may be explained by the unbalanced distribution of domains in the NTCIR-2 collection.
Most importantly, \textit{out-domain} queries on full indexing (i.e.~$T$+$A$+$K$) is the only configuration in which no significant gains in retrieval effectiveness are achieved.
This last experiment shows that expanding documents using existing keyphrase generation models may be ineffective in the absence of in-domain training data, and stresses the need of domain adaptation for keyphrase generation.

\begin{table}[ht!]
    \centering
    \setlength{\tabcolsep}{4pt}
    \resizebox{.485\textwidth}{!}{%
    \begin{tabular}{l|cc|cc}
        \toprule
        \multirow{2}{*}{\textbf{Model}} & \multicolumn{2}{c|}{$T$+$A$} & \multicolumn{2}{c}{$T$+$A$+$K$}\\
        
        
        ~ & \textbf{I}~{\normalsize(32.70)} & \textbf{O}~{\normalsize(30.99)} & \textbf{I}~{\normalsize(36.18)} & \textbf{O}~{\normalsize(33.93)} \\
        \midrule

        s2s+copy &
        \sign{\textcolor{good}{\textbf{35.40}}} & \sign{\textcolor{good}{\textbf{32.96}}} & \sign{\textcolor{good}{\textbf{38.13}}} & \textcolor{good}{\textbf{34.55}} \\
    
        s2s+corr &
        \textcolor{good}{33.49} & \textcolor{good}{32.92} & \textcolor{good}{37.13} & \textcolor{good}{34.25} \\
        
        mp-rank &
        \textcolor{good}{32.73} & \textcolor{good}{31.71} & \textcolor{good}{36.74} & \textcolor{bad}{33.26} \\
        
        \bottomrule
    \end{tabular}
    }
    \caption{MAP scores for BM25+RM3 on \textit{in-domain}~(I) and \textit{out-domain}~(O) queries. $\dagger$~indicates significance over w/o keyphrases whose scores are in parentheses.}
    \label{tab:per_domain}
\end{table}

\section{Conclusion}

We presented the first study of the usefulness of keyphrase generation for scientific document retrieval.
Our results show that keyphrases can significantly improve retrieval effectiveness, and also highlight the importance of evaluating keyphrase generation models from an extrinsic perspective.
Other retrieval tasks may also benefit from using keyphrase information and we expect our results to serve as a basis for further improvements.

\section*{Acknowledgements}

We thank the anonymous reviewers for their valuable comments.
This work was supported by the IKEBANA project (grant of Atlanstic 2020) and the French National Research Agency (ANR) through the DELICES project (ANR-19-CE38-0005-01).

\bibliographystyle{acl_natbib}
\bibliography{references}

\newpage
~
\newpage







\colorlet{color1}{blue!70!black}
\colorlet{color2}{purple}

\appendix

\section{Supplementary material}

\subsection{Related Work}

\subsubsection*{Keyphrase extraction and generation}


Identifying keyphrases for a given document is a long standing task in NLP.
Earlier work typically involves two steps: 1)~extracting keyphrase candidates; and 2)~ranking those candidates by importance.
Models mainly differ in how they do the latter, commonly used techniques being supervised learning~\cite{Witten:1999:KPA:313238.313437,Turney:2003:CKE:1630659.1630724,Nguyen:2007:KES:1780653.1780707,Jiang:2009:RAK:1571941.1572113} and graph-based methods~\cite{mihalcea-tarau:2004:EMNLP,wan-xiao:2008:PAPERS,bougouin-boudin-daille:2013:IJCNLP,florescu-caragea:2017:Long}.
These models are, however, inherently limited in the sense that they can only output keyphrases that appear in the text.
To allow the prediction of keyphrases describing implicit topics or using different wordings, previous work relied on external resources like controlled vocabularies~\cite{4119142,bougouin-etal-2016-keyphrase}, while recent attempts leveraged neural generative models~\cite{meng-etal-2017-deep,chen-etal-2018-keyphrase,zhao-zhang-2019-incorporating}.

\subsubsection*{Biomedical indexing}

Also related to our work is the research done on biomedical semantic indexing using MeSH\footnote{\url{https://www.nlm.nih.gov/mesh/}}, a hierarchically-organized controlled vocabulary. 
Automated methods for assigning MeSH terms make use of all sorts of techniques, such as pattern matching~\cite{aronson2004nlm} or learning to rank~\cite{liu2015meshlabeler,peng2016deepmesh}.

%


\subsubsection*{Document expansion}

Our work is similar in nature to previous research on document expansion~\cite{tao-etal-2006-language,Efron:2012:IRS:2348283.2348405}, and is closely related to recent work on document expansion using automatically generated queries~\cite{nogueira2019document}.

\subsection{Parameters}

Table~\ref{tab:param} displays the model parameters we use for seq2seq+copy and seq2seq+corr.

\begin{table}[!ht]
    \centering
    \begin{tabular}{l|c}
    \toprule
        \textbf{Parameter} & \textbf{Value} \\
    \midrule
        Network & bi-GRU \\
        Vocabulary size & 50K \\
        Word embedding size & 150 \\
        Hidden layer & 2 \\
        Hidden layer size  & 300 \\
        Optimizer & Adam \\
        Initial learning rate  & $10^{-4}$ \\
        Gradient clipping & 0.1 \\
        Dropout & 0.5 \\
        Beam depth & 6 \\
        Beam size & 200 \\
    \bottomrule
    \end{tabular}
    \caption{Model Parameters.}
    \label{tab:param}
\end{table}




%



Table~\ref{tab:in-out-split} presents the research fields used for dividing queries into two sets.

\begin{table}[ht!]
    \centering
    \begin{tabular}{l|ccc}
        \toprule
        \textbf{Research field} & \textbf{In} & \textbf{Out} \\
        \midrule
        Electricity, information and control & \checkmark & - \\
        Chemistry & \checkmark & - \\
        Architecture, civil engineering 
            & - & \checkmark \\
        Biology and agriculture & - & \checkmark \\
        Science  & \checkmark & - \\
        Engineering  & \checkmark & - \\
        Medicine and dentistry  & - & \checkmark \\
        Cultural and social science & - & \checkmark \\
        \midrule
        \multicolumn{1}{r|}{\textbf{\# of queries}}
         & 27 & 22 \\
        \bottomrule
    \end{tabular}
    \caption{Research fields for in- and out-domain queries.}
    \label{tab:in-out-split}
\end{table}


\subsection{Example}

An example of document along with automatically generated keyphrases is shown in Table~\ref{tab:example}.

\begin{table*}[htb!]
    \centering
    \begin{tabular}{r p{0.8\textwidth}}
    \toprule
    title & \textcolor{color1}{Grammatical Inference} for Concept Acquisition from Documents. \\
    \midrule
    abstract & The purpose of this study is to acquire knowledge from large scale natural language documents. There are two types of knowledge in the documents. One is explicitly represented knowledge which is acquired using natural language processing. The other is implicit constrain. In this paper, how to acquire implicit constraint using \textcolor{color1}{grammatical inference} from the documents is described. We propose a \textcolor{color1}{grammatical inference} system which uses inference rules based on logic, and show that the system can learn easy pattern of character lists. We also discuss its application to \textcolor{color2}{knowledge acquisition} from real documents. \\
    \midrule
    gold & \textcolor{color1}{grammatical inference} // \textcolor{color2}{knowledge acquisition} // logic \squiggly{progamming} // concept learning \\
    \midrule
    s2s+copy & \textcolor{color1}{grammatical inference} // knowledge // grammatical // knowledge representation // natural language processing \\
    \midrule
    s2s+corr & \textcolor{color1}{grammatical inference} // \textcolor{color2}{knowledge acquisition} // concept acquisition // inference rules // natural language processing \\
    \midrule
    mp-rank & \textcolor{color1}{grammatical inference} // documents // knowledge // large scale // concept acquisition \\
    \bottomrule
    \end{tabular}
    \caption{Example document (id: gakkai-e-0001014453 from ntc2-e1g) with author keywords (gold) and automatically generated keyphrases. We note a typo in the gold annotation (progamming).}
    \label{tab:example}
\end{table*}





\end{document}